\documentclass[journal, doublecolumn]{IEEEtran}

\usepackage{blindtext, graphicx}
\usepackage{amsmath}
\usepackage{amssymb}
\usepackage{subfigure}
\usepackage{cite}
\usepackage{amsxtra}
\usepackage{amsfonts}
\usepackage{graphicx}
\usepackage{multirow} 
\usepackage{listings}
\usepackage{amsmath}
\usepackage{caption}
\usepackage{blindtext}
\usepackage{rotating}
\usepackage{multirow}
\usepackage[table,xcdraw]{xcolor}
\usepackage{verbatim} 

\usepackage{colortbl}

\usepackage{hhline}
\usepackage[T1]{fontenc}
\usepackage[utf8]{inputenc}
\usepackage{tabularx,ragged2e,booktabs,caption}
\newcolumntype{C}[1]{>{\Centering}m{#1}}

\usepackage{chngcntr}
\usepackage{lipsum}
\usepackage[font=small,labelfont=bf]{caption}
\newenvironment{Table}
   {\par\bigskip\noindent\minipage{\columnwidth}\centering}
   {\endminipage\par\bigskip}

\ifCLASSINFOpdf
\else
\fi
\begin{document}

%
\title{5G Coverage, Prediction, and Trial Measurements}
%
%
%

\author{%
\begin{tabular}{c} Tristan Curry \\ Macquarie University  \\ tristan.curry@students.mq.edu.au \end{tabular} \and
\begin{tabular}{c} Robert Abbas \\ Macquarie University \\ robert.abbas@mq.edu.au \end{tabular} \and 
}

\maketitle

\begin{abstract}
When planning a 5G network in the sub-6GHz bands, similar cell planning techniques to LTE can be applied. Looking at the Australian environment, the n78 band (3.3-3.8GHz TDD) is approximately 1GHz higher than the 2.6GHz band used in existing LTE networks. As a result, the coverage footprint can be similar, and therefore co-locating 5G NR (New Radio) on existing LTE base stations is a common strategy for initial network rollout. Any difference in coverage can be compensated by beamforming gain, less downtilting, or increasing the gNodeB's transmit power. This paper presents an initial link budget for data services, provides a coverage prediction, and measurements for a 5G NR NSA (Non Stand Alone) trial radiating at 3.5GHz with 60 MHz bandwidth. The coverage prediction is generated using RF planning tool Atoll, which is then compared to coverage measurements from the trial. These findings can be used to help plan a future 5G network in the Sydney metro area or similar environment.

\end{abstract}

\IEEEpeerreviewmaketitle

\section{Introduction}
\indent 5G has support for spectrum ranging from existing microwave bands as low as 600MHz up to mmWave bands (>30GHz band). With higher carrier frequencies comes increased bandwidth but higher penetration losses, so planning considerations such as inter-site distance and base station architectures used will change accordingly. In November 2018, the Australian Communications and Media Authority (ACMA) auctioned off spectrum in the 3.6GHz band  \cite{Ref5} in anticipation of 5G networks using this band. The recommendation to use this band is mirrored by the 3GPP \cite{Ref2}.  It is anticipated that most Mobile Network Operators (MNO) in Australia will use the n78 band in NR frequency range 1 (FR1) for initial 5G network roll-out so therefore this trial was undertaken as a study of how 5G NR behaves in these bands. 
\newline
\indent To determine the minimum received signal level at the user equipment (UE) required at cell edge, a link budget was applied. This minimum receive level was then used to create a coverage prediction, with contour lines corresponding to different NR Reference Signal Received Power (NRSRP) levels. The coverage prediction was generated using the 5G NR Atoll RF simulation suite and the CrossWave propagation model. These were applied to high resolution raster (2m pixels) digital terrain maps (DTM) and clutter data, and 3D vector building and vegetation data in the trial area. This coverage prediction was then compared to measured data and the accuracy of the predictions was determined. This information can be used to plan contiguous 5G coverage in the Sydney metro area. 
\newline
\indent The focus of this paper is to present a 5G NR link budget and corresponding prediction, and compare this against measured NRSRP and throughput data from a live 5G NR site to verify the accuracy of the link budget and prediction models for planning a 5G network.

\begin{figure*}[h]
\centering
\captionsetup{justification=centering}
\includegraphics[height=6.5cm, width=14.5cm]{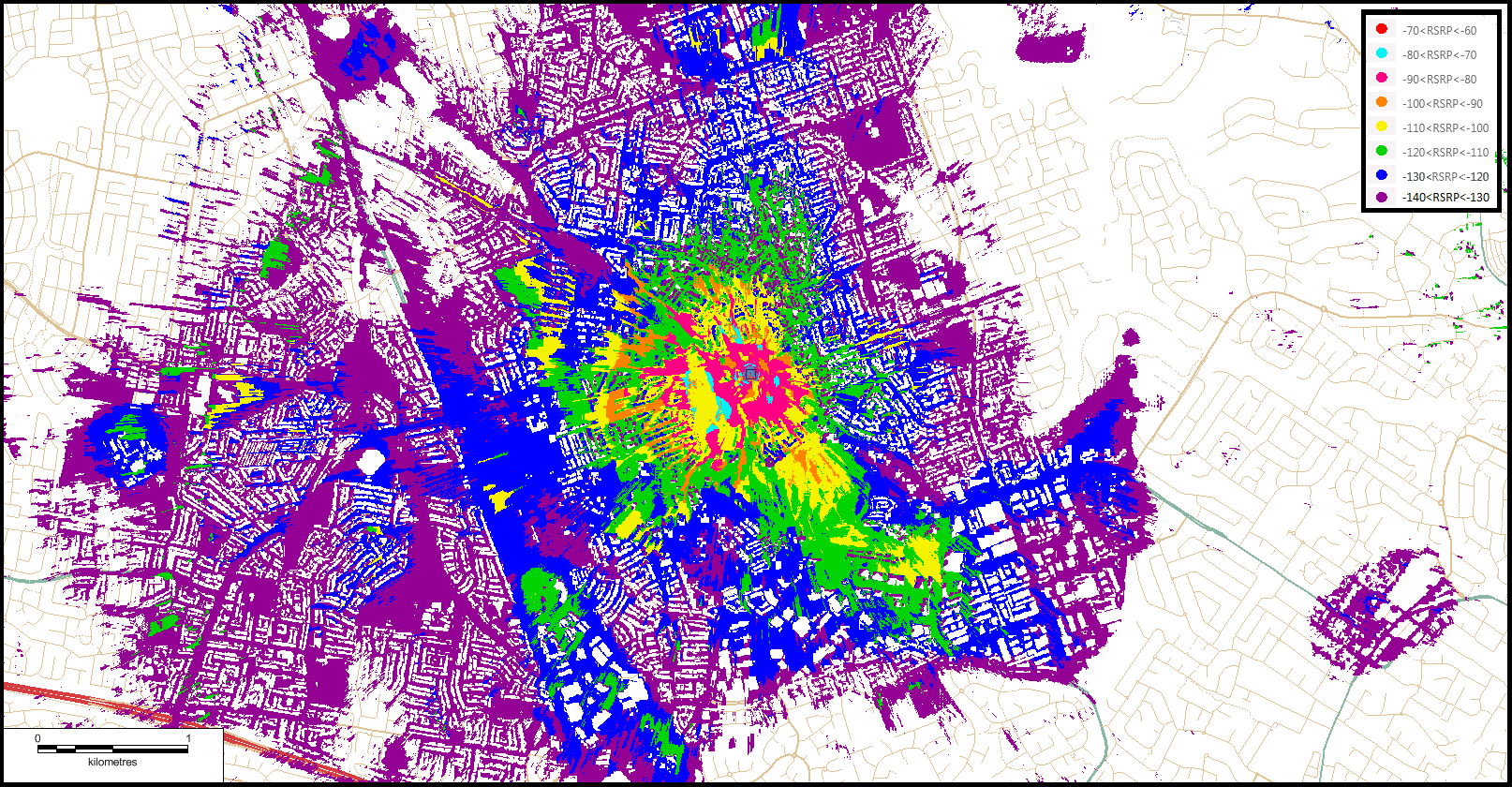}
\caption{NR Coverage Prediction using the Atoll Network Planning tool and CrossWave Propagation Model }
\end{figure*}

\section{Background}
\indent In this section we detail some of the relevant work that has been done in both academia and industry. We note that there are very few live trial results which have been published in the n78 band for mobile networks.
\newline 
\indent In \cite{Ref3} the authors present a paper comparing link budgets for the 2.1 GHz band for LTE and the 3.5 GHz band for NR. They conclude that in a direct comparison the 3.5 GHz band has a number of extra losses in both indoor and outdoor environments, however when using beam-forming the associated gain can overcome these losses. They deduced that coverage from the 3.5GHz band was better in outdoor environments, and indoor in older buildings, but only 0.9dB different in buildings built in a modern style.
\newline
\indent In \cite{Ref6}, the authors provide an analysis of TCP performance over 5G NR. They explore the effect of the NSA architecture on overall TCP throughput. They deduce that due to the differing latency performance of NR and anchor LTE layers, there is degradation in throughput in spite of any increased spectrum bandwidth. They also deduce that the nature of LTE and NR mean that TCP is susceptible to head-of-line blocking by regular data traffic. This effect de-prioritises TCP ACKs and reduces overall throughput by reducing the congestion window. When testing 5G NR using TCP based methods such as Ookla Speedtests, these effects should be taken into account.
\newline
\indent There have been 5G trials conducted around the world by operators and vendors, Telstra and Ericsson in Australia demonstrated a 5G call using an over-the-top service on an NSA 5G NR RAN and EPC core using the n78 band at the Gold Coast Commonwealth Games in April 2018 \cite{Ref7}. In California, Nokia and Keysight have demonstrated the capabilities of Keysight's Nemo network measurement equipment to help optimise and verify the performance of Nokia's 5G NR RAN equipment \cite{Ref8}.
\newline
\indent We base our understanding and research of NR on the 3GPP Technical Reports (TR) and Technical Specifications (TS) documents. The current release (3GPP release 15) details the specification for 5G NR \cite{Ref2,Ref1} and outlines the requirements for NR standalone (SA) and non-standalone (NSA). Due to equipment availability and maturity, initial deployment will have the NR NSA architecture, utilising LTE on E-UTRA band 1 (2.1GHz FDD) for the anchor layer. This anchor layer provides signalling to allow the UE to set up an NR bearer when in dual connectivity mode and thus connect to the gNodeB. This arhcitecture can be seen in Fig. 2.

\begin{figure}[h]
\captionsetup{justification=centering}
\includegraphics[height=4cm, width=8.5cm]{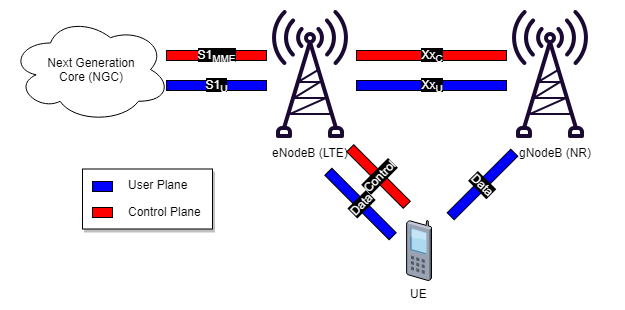}
\caption{5G NR Non-Standalone (NSA) Architecture}
\label{fig:fig3}
\end{figure}

\section{Link Budget and Coverage Prediction}

\indent Industry best practice for cell planning starts with determining an appropriate receive level at the UE. This level in conjunction with coverage prediction tools will determine where the cell edge is. A link budget provides the basis for what service level the network is designed for and outlines the gains and losses in the mobile communications channel. This information allows a final received power level to be calculated which corresponds to the desired service level. We based the channel parameters on the 3GPP NR channel model \cite{Ref2}, and have chosen system bandwidth, MIMO schemes, and heights based on what NR RAN equipment was available at the time of the trial. A summarised 5G NR link budget is shown in table I. This link budget suggests we will achieve 200Mbps at a UE received NRSRP level of -90.62 dBm. 
\newline
\indent Based on 3GPP recommendations \cite{Ref2}, the 3.5GHz band is preferred for initial NSA 5G networks and the ACMA has followed this recommendation in Australia by auctioning off more spectrum in the 3.4-3.6GHz bands for mobile communications use \cite{Ref5}. We have therefore used 60MHz bandwidth centered at 3.5GHz in licensed spectrum for this trial.

\subsection{gNodeB Physical Configuration}

\begin{Table}
\captionsetup{justification=centering}
\captionof{table}{5G NR Link Budget}
\includegraphics[width=8cm]{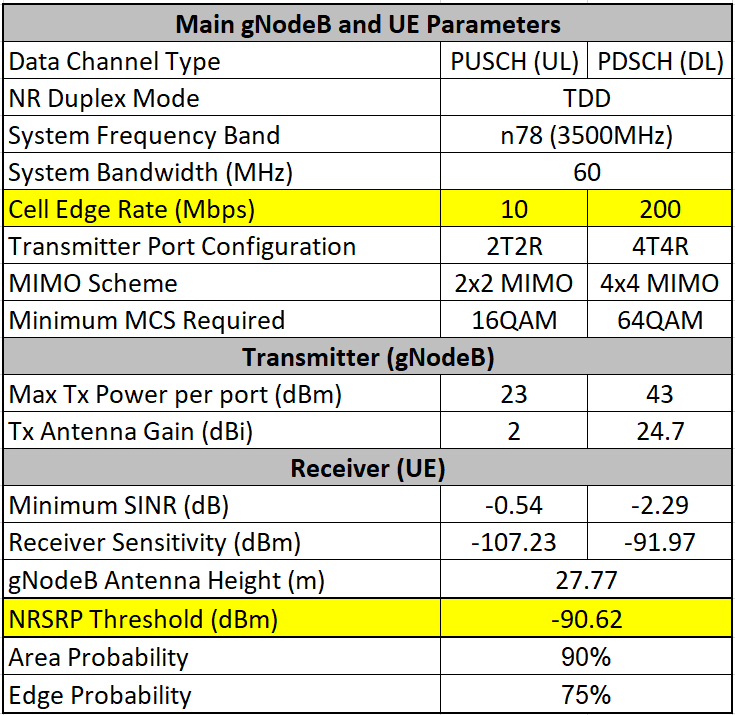}
\end{Table}

\indent The gNodeB was built in a typical suburban area in Sydney. The physical site parameters were based on the 3GPP channel model seen in \cite{Ref2} with the antenna center line (ACL) height raised up to 27.77 m above the ground due to space constraints on the headframe, and 3$^\circ$ degrees electrical downtilt to reduce backlobe interference between sectors and increase coverage. As there were no neighbouring gNodeB's, 0$^\circ$ mechanical downtilt was used to maximise coverage. 
\newline
\indent The antennas used were passive antennas with 8 antenna elements. In the NR specification coverage is provided by Synchronisation Signal Block (SSB) beams that sweep across the width of the antennas' coverage envelope. The antenna used swept through 8 SSB beams in a round-robin manner from index 0 to 7. 
\newline
\indent The terrain in the area varied in elevation from 50.69 m to 125.65 m above sea level, with clutter primarily being comprised of small shopping precincts, trees, housing developments, and industrial areas.

\subsection{Coverage Prediction}

\indent Based off our link budget minimum receive level at the UE we have generated a coverage prediction in Forsk Atoll using the CrossWave propagation model. CrossWave is a pseudo-raytracing model which emulates raytracing when used in conjunction with high enough resolution raster or vector clutter data. We used 2m resolution raster clutter data with outdoor to indoor propagation enabled. The coverage prediction for the site can be seen in Fig. 1. 
\newline 
\indent We note that we expect to achieve 200Mbps DL throughput in the orange area (-100dBm < NRSRP < -90dBm)  based on the link budget.

\section{Measurements and Results }

\subsection{Testing Methodology}

\indent We conducted 2 tests during this trial. The first was a drive test using a Rohde \& Schwarz NR scanner connected to a PCTEL omni-directional antenna magnetically mounted to the roof of the car. This antenna has an average of 3dBi gain in the range 698-3800MHz. The scanner logged data points in Single Input, Single Output (SISO) mode over more than 30km in surrounding streets. The second test was a stationary throughput test using a 5G UE (User Equipment) set up in the window in the back of the car. The UE utilised 2x2 Multiple Input, Multiple Output (MIMO) and speed tests were conducted using a TCP based speedtest service on a smartphone connected to the UE over 5 GHz WiFi. We collected 32 samples in the one location. 
\newline
\indent From the prediction, we determined the drive test area and drove both near the base station and towards the cell edge.
\newline
\indent Key data collected with the scanner included:
\begin{itemize}
    \item Timestamp (ms).
    \item UE GPS (Global Positioning System) Coordinates.
    \item SSB Beam index.
    \item NRSRP (dBm) per beam.
    \item NRSRQ (NR Reference Signal Received Quality) (dB) per beam.
\end{itemize}

\indent Key data collected with the 5G UE included:
\begin{itemize}
    \item Downlink Throughput (Mbps).
    \item Uplink Throughput (Mbps).
    \item Latency (ms).
    \item NRSRP (dBm).
\end{itemize}

\subsection{Measured Drive Test Results} 

\indent The results were obtained over 2 days. Measured NRSRP values along the drive test route can be seen in Fig. 3 (a).

\begin{figure}[h]
\centering
\captionsetup{justification=centering}
\includegraphics[height=5cm, width=6.5cm]{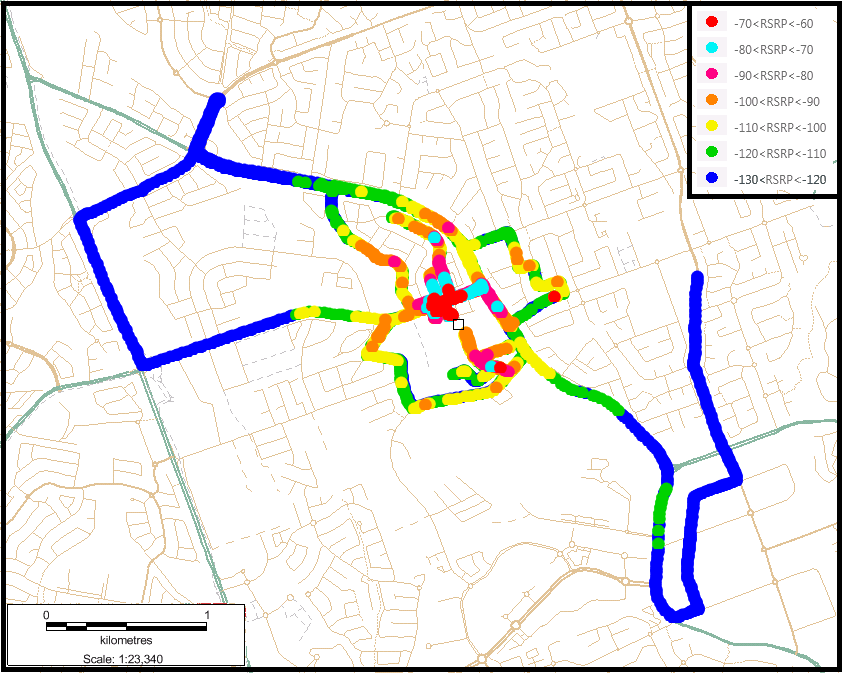}
\includegraphics[height=5cm, width=6.5cm]{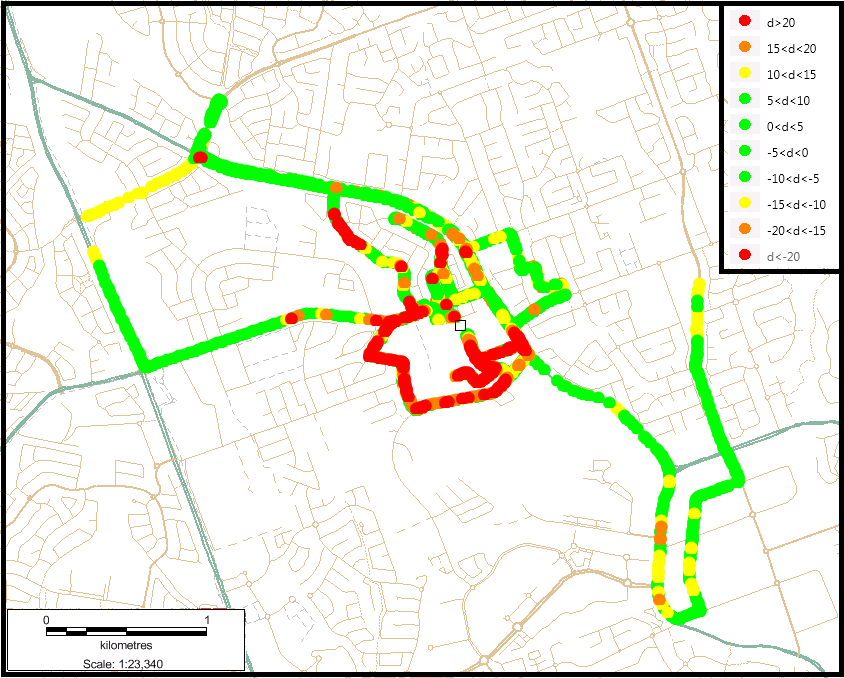}
\caption{Coverage prediction route as per measurement route: (a) above, Scanner NRSRP measurements, (b) below, delta of measured to predicted NRSRP values.}
\end{figure}

\indent NRSRP level with respect to distance is summarized in Fig. 4. The NRSRP fluctuations close to the gNodeB can be attributed to fast fading as a result of higher multipath fluctuation as the UE moves in the urban canyon areas near the site. This fluctuation can be seen to be above 20dB in areas close to the site, but this variation reduces further from the site as slow fading becomes the dominant fading. 
\begin{figure}[h]
\captionsetup{justification=centering}
\includegraphics[height=3.5cm, width=8.5cm]{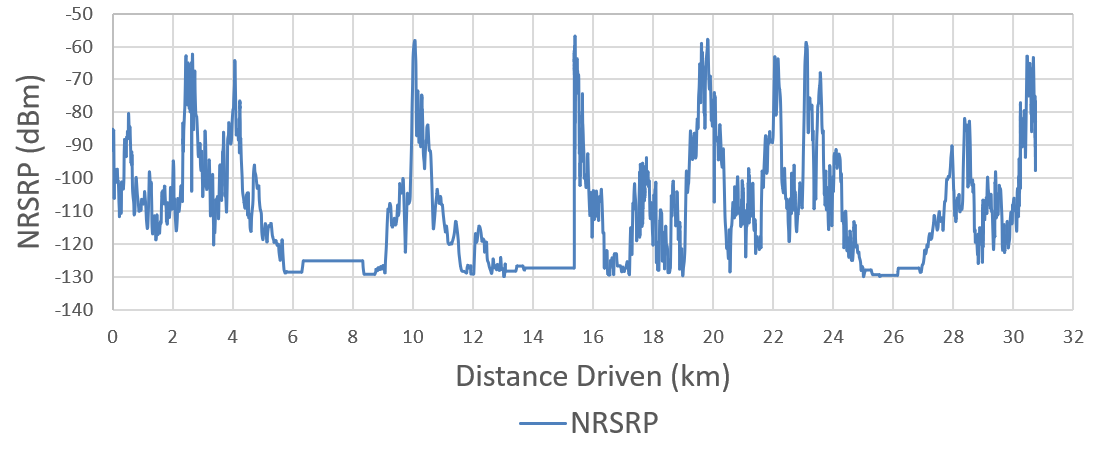}
\caption{Drive Test NRSRP Measurements}
\label{fig:fig4}
\end{figure}

\indent To see the effect of slow fading only, we apply a general form of Lees method \cite{Ref9}. To compute the local averages of the signal envelope we used the recommendation from Lee for an averaging window \(2L = 40\lambda\) which corresponds to a distance \(D = 342.86 cm\), minimum number of samples per average window \(N = 36\), and a distance between samples \(d = 0.8\lambda = 6.86cm\) \cite{Ref9} with a 3500MHz carrier (\(f = 3500MHz, \lambda = 8.57cm\)). These values are chosen to ensure when the moving average of the signal envelope is calculated, the fast fading component becomes unity, and therefore does not affect the slow fading component. The normalised signal envelope after this process shows only the slow fading of the signal and can be seen in Fig. 5.  This data can be used to tune the CrossWave model using the built in tool which utilises curve fitting to adjust parameters specific to the model to the measured and normalised signal envelope.

\begin{figure}[h]
\captionsetup{justification=centering}
\includegraphics[height=3.5cm, width=8.5cm]{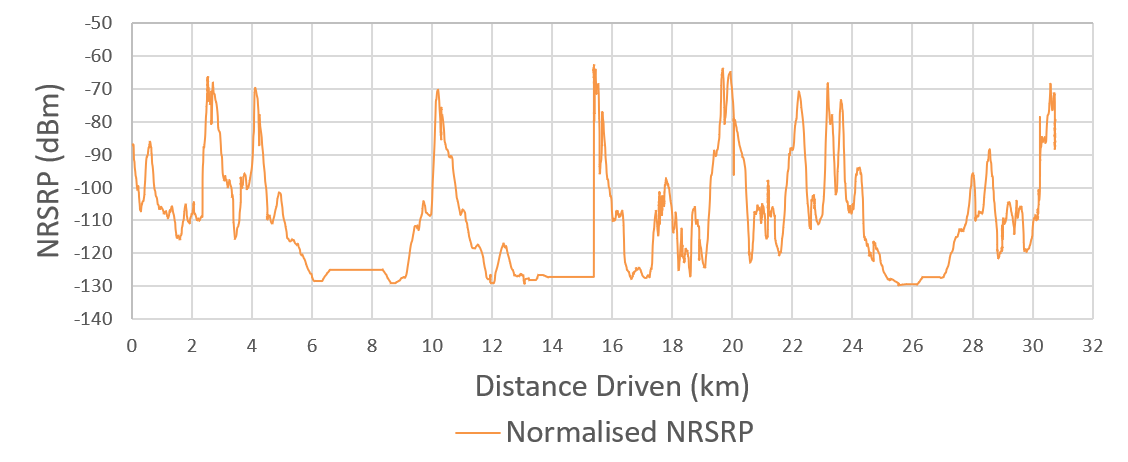}
\caption{Normalised NRSRP Measurements over Drive Test Distance}
\label{fig:fig5}
\end{figure}

\indent  Fig. 6 shows the NRSRQ and SINR values for each beam from sector 1 at a location in the drive test route. We can see that NRSRQ and SINR have varying values for each of the 8 beams. Beam index 6 was directed at the measurement location and had the highest NRSRQ and SINR values, and beam index 2 had a dominant reflected path which gave a high NRSRQ and SINR value at the measured location. Adjacent beams were not pointing directly at the measurement location and therefore less SSB power was received at the scanner. This corresponds to a reduction in signal strength, and therefore NRSRQ and SINR. The theoretical relationship between NRSRQ and SINR is:
\[ SINR = \frac{1}{\frac{1}{n \times NRSRQ} - x}\]
Where: \(n = \) number of subcarriers, and \(x = \) per antenna subcarrier activity factor.
\newline 
\indent From this relationship it can be seen that NRSRQ does not fluctuate as highly as SINR for this range of values as they lie in the non-linear region in the SINR to NRSRQ graph. 

\begin{figure}[h]
\captionsetup{justification=centering}
\includegraphics[width=8.8cm]{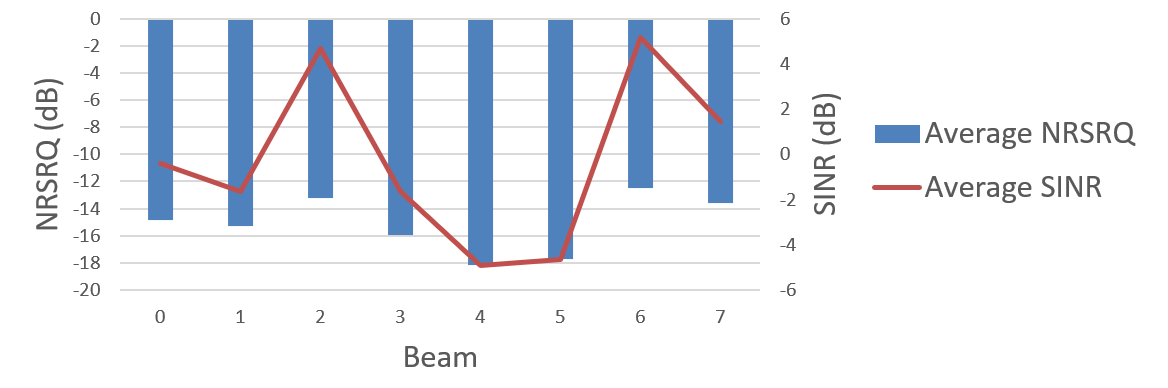}
\caption{NRSRQ and SINR values per beam.}
\label{fig:fig2}
\end{figure}

\subsection{Measured Throughput Test Results} 

\indent Fig. 7 represents an average throughput comparison between downlink, and uplink, and latency at an NRSRP of -91dBm {$\scriptstyle\pm$} 1dB. The sample size \(n = 32 > 30\) allows us to assume normality from the Central Limit Theorem (CLT). 

\begin{figure}[h]
\captionsetup{justification=centering}
\includegraphics[width=8.5cm]{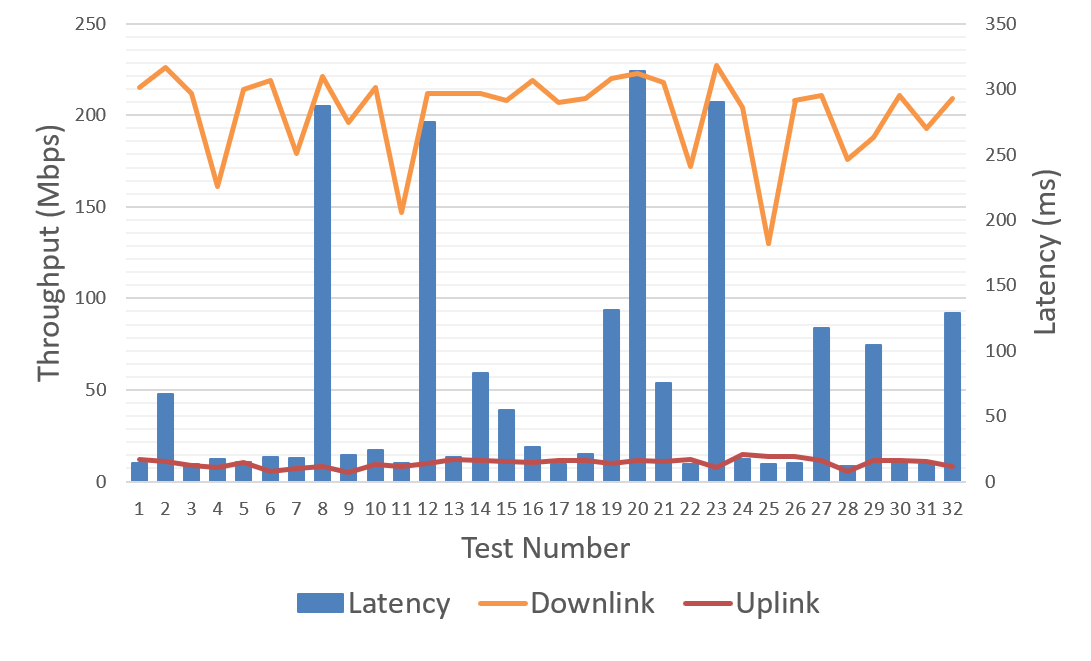}
\caption{Speedtest and Latency Results at -91dBm {$\scriptstyle\pm$} 1dB NRSRP}
\label{fig:fig1}
\end{figure}

\indent Summary statistics can be seen in Table II. These results align with what we expected from the link budget, with a throughput of 200Mbps at approximately -91dBm. The high variance in latency is due to outliers reaching 300ms. These skew the average and can be seen when looking at the median value of only 19.5ms. We believe this is due to IP routing issues in the internet causing packet loss and therefore high roundtrip times (RTT).

\begin{Table}
\captionsetup{justification=centering}
\captionof{table}{Speedtest Summary Statistics}
\includegraphics[width=8cm]{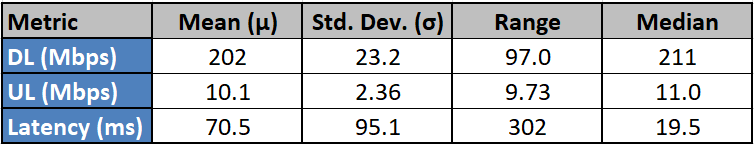}
\end{Table}

\section{Conclusion and Future Work}

\indent Based on our analysis and application of LTE cell planning techniques, we conclude they can be applied in initial 5G NR network rollout in the n78 band. This trial demonstrated that 200Mbps was possible at -91dBm {$\scriptstyle\pm$} 1dB NRSRP using a link budget analysis and trial measurements. The coverage prediction and measured values after using Lee's method aligned well and give confidence in the prediction for further network planning in similar environments. In future we suggest to apply the same analysis to more locations and environments, and perform speedtests at more NRSRP levels and locations with an FTP speedtest methodology that doesn't have the variability associated with over-the-top speedtest services. With the focus on fixed wireless access (FWA) in the United States and Australian markets, we suggest research into benchmarking the penetration of the n78 band through different clutter types and materials found in the Australian environment to generate more accurate link budgets. We recommend testing the performance of horizontal network handover in a mobility cluster of sites, as this will also influence the density of sites when planning a 5G network.

\bibliographystyle{elsarticle-num}

\end{document}